\documentclass[twocolumn,prd,aps,floats,nofootinbib,tightenlines,showpacs,floatfix]{revtex4-1}
\pdfoutput=1
\usepackage{graphicx}
\usepackage{amsmath}
\usepackage{amsfonts}
\usepackage{amssymb,amsthm}
\usepackage{mathrsfs}
\usepackage{epsfig}
\usepackage{color}
\begin{document}
\title{Thermal photons in QGP and non-ideal effects}
\author{Jitesh R. Bhatt}
\email{jeet@prl.res.in}
\author{Hiranmaya Mishra}
\email{hm@prl.res.in}
\author{V. Sreekanth}
\email{skv@prl.res.in}
\affiliation{Physical Research Laboratory, Ahmedabad
380009, India}
\begin{abstract}
 We investigate the thermal photon production-rates using one dimensional boost-invariant second order relativistic 
hydrodynamics to find  proper time evolution of the energy density and the temperature. The effect of bulk-viscosity and
non-ideal equation of state are taken into account in a manner consistent with recent lattice QCD estimates. 
It is shown that the \textit{non-ideal} gas equation of state i.e $\varepsilon-3\,P\,\neq 0$ behaviour of the expanding plasma,
which is important near the phase-transition point, can significantly slow down the hydrodynamic expansion and 
thereby increase the photon production-rates. Inclusion of the bulk viscosity may also have similar effect on the hydrodynamic evolution. 
However the effect of bulk viscosity is shown to be significantly lower than the \textit{non-ideal} gas equation of state.  
We also analyze the interesting phenomenon of bulk viscosity induced cavitation making the hydrodynamical description
invalid. We include the viscous corrections to the distribution functions while calculating the photon spectra. 
It is shown that ignoring the cavitation phenomenon can lead to erroneous estimation of the photon flux.

\end{abstract}
\maketitle
\section{INTRODUCTION}

Thermal photons emitted from the hot fireball created in relativistic heavy-ion
collisions is a promising tool for providing a signature of quark-gluon plasma
\cite{kapu91,baier92,Ruus92,thoma95,traxler95,arnoldJHEP} 
(see \cite{Alam01R,Peitz-Thoma:02R,Gale:03R} for recent reviews). 
Since they participate only in electromagnetic 
interactions, they 
have a larger mean free path compared to the transverse size of the hot and dense matter created 
in nuclear collisions 
\cite{kapusta}. 
Therefore  these photons were proposed to verify the existence of the QGP phase 
\cite{Feinberg76,Shyryak78}.
Spectra of thermal photons depend upon the
fireball temperature and they can be calculated from the scattering cross-section of
the processes like $q\bar q\rightarrow g\gamma$, \textit{bremsstrahlung} etc. Time evolution
of the temperature can be calculated using hydrodynamics with appropriate
initial conditions. 
Thus the spectra depend upon the equation of state (EoS) of the medium and they 
may be useful in finding a signature of the quark-gluon plasma\cite{dk99:EPJC,thoma07,dkdir08,Liu09}. 
Recently thermal photons are proposed as a tool to measure the shear viscosity
of the strongly interacting matter produced in the collisions\cite{skv-photon,dusling09}.

 Understanding the shear viscosity of QGP is one of the most
intriguing aspects of the experiments at Relativistic Heavy Ion Collider (RHIC).
Analysis of the experimental data collected from RHIC shows
that the strongly coupled matter produced in the collisions is not too much above
the phase transition temperature $T_c$ and it may have extremely small value of shear viscosity
$\eta$. The ratio of the shear viscosity
$\eta$ to the entropy density $s$ i.e. $\eta/s$ is around $1/4\pi$ which is the smallest for
any known liquid in the nature\cite{Schf-Teaney-09R}. 
In fact the arguments based on AdS-CFT suggest that 
the values of $\eta/s$ can not become lower than $1/4\pi$. This is now 
known as  Kovtun-Son-Starinets or 'KSS- bound' \cite{kss05}. Thus the quark-gluon plasma
produced in RHIC experiments is believed to be in a form of the most perfect liquid\cite{Hirano:2005wx}.
No wonder ideal hydrodynamic appears to be the best description of such matter as suggested by
comparison between the experimental data\cite{RHIC} and the calculations done using
second-order relativistic hydrodynamics
\cite{BR07,rom:PRL,DT08,hs08,LR08,Molnar08,Song:2008hj,LR09}.

However there remain uncertainties in understanding the application and validity
of the hydrodynamical procedure in relativistic heavy-ion collision experiments. 
It is only very recently realized that the effect of bulk viscosity can bring complications
in the hydrodynamical description of the heavy-ion collisions. 
Generally it was believed that the bulk viscosity, $\zeta$ does not play a significant
role in the hydrodynamics of relativistic heavy-ion collisions. 
It was argued that that since $\zeta$ scaled like  $\varepsilon -3P$ 
at very high energy the bulk viscosity may not play a significant role because
the matter might be following the ideal gas type equation of state\cite{wein}. But during its course
of expansion the fireball temperature can  approach values close to $T_c$. 
Recent lattice QCD results show that the quark-gluon matter do not satisfy ideal EoS near $T_c$ 
and the ratio $\zeta/s$ show a strong peak around $T_c$ \cite{Bazavov:2009,Meyer:2008}. 
The bulk viscosity contribution in this regime can be much larger than that of the the shear viscosity.
Recently the role of bulk viscosity in heating and expansion of the fireball was analyzed
using one dimensional hydrodynamics\cite{fms08}. 
Another complication that bulk viscosity brings in hydrodynamics of
heavy-ion collisions is phenomenon of cavitation\cite{kr:2010cv}. Cavitation arises
when the fluid pressure becomes smaller than the vapour pressure. Since the
bulk viscosity (and also shear viscosity) contributes to the pressure gradient
with a negative contribution, it may be possible for the effective fluid pressure
to become zero. Once the cavitation sets in, the hydrodynamical description
breaks down. It was shown in Ref.\cite{kr:2010cv} that cavitation may happen in
RHIC experiments when the effect of bulk viscosity is included in manner consistent
with the lattice results. It was shown that the cavitation may significantly
reduce the time of hydrodynamical evolution. 

Keeping the above discussion in mind, we aim to study the effect of bulk viscosity and 
cavitation on the thermal photon production in heavy ion collisions. 
As far as we know no such study exists in the literature. Furthermore, the calculations for the photon 
production rates in the absence of viscous effects, are done using thermal distribution function of the particle species 
(e.g., quark, anti-quark etc.)\cite{Peitz-Thoma:02R}. However, it is well known that viscous effects can lead to the modification of 
the thermal distribution functions\cite{deGroot}. This may have observational effect on the photon spectra\cite{dusling09}. 
In this work we incorporate the viscous modification in the distribution function arising due to bulk and shear viscosities. 
Finite $\zeta$ effect can either 
significantly reduce the time for the hydrodynamical evolution (by onset of cavitation)
or  it can increase the time by which the system reaches $T_c$. Moreover the \textit{non-ideal}
gas EoS can also significantly influence the hydrodynamics. 
In what follows, we use equations of relativistic second order hydrodynamics to
incorporate the effects of finite viscosity. We take the value of $\zeta/s$ same
as that in Ref.\cite{kr:2010cv} and keep $\eta/s=1/4\pi$. Further we use one dimensional
boost invariant hydrodynamics in the same spirit as in Refs.\cite{fms08,kr:2010cv}.
One of the limitations of this approach is that the effects of transverse flow cannot
be incorporated. As the boost-invariant hydrodynamics is known to lead to under-estimation
of the effects of bulk viscosity\cite{fms08}, we believe that our study of the photon
spectra will provide a conservative estimate of the effect. However it should also be noted that the effect of 
transverse flow could remain small as cavitation can restrict the time for hydrodynamical evolution. 

\section{FORMALISM}
\subsection{Viscous Hydrodynamics}
We represent the energy momentum tensor of the dissipative QGP formed in high energy nuclear collisions as
\begin{equation}
T^{\mu\nu}=\varepsilon \, u^\mu\,u^\nu\, - P\, \Delta^{\mu\nu} + \Pi^{\mu\nu}
\label{Tmunu}
\end{equation}
where $\varepsilon$, $P$ and $u^\mu$ are the energy density, pressure and four velocity of the fluid element respectively. 
The operator $\Delta^{\mu\nu}~=~g^{\mu\nu} - u^\mu\,u^\nu$ acts as a projection perpendicular to four velocity. 
The viscous contributions to $T^{\mu\nu}$ are represented by 
\begin{equation}
 \Pi^{\mu\nu} = \pi^{\mu\nu}-\,\Delta^{\mu\nu}\, \Pi
\end{equation}
where $\pi^{\mu\nu}$, the traceless part of $\Pi^{\mu\nu}$; gives the contribution of shear viscosity and $\Pi$ gives the 
bulk viscosity contribution. The corresponding hydrodynamics equations are given by,
\begin{eqnarray}
D \varepsilon + (\varepsilon+P)\, \Theta-\Pi^{\mu\nu}\nabla_{(\mu}\,u_{\nu)}=0\label{edot}\\
(\varepsilon+P)\,Du^{\alpha}-\nabla^{\alpha}P\,+\,\Delta_{\alpha\nu}\,\partial_\mu \Pi^{\mu\nu}=0
\end{eqnarray}
where $D\equiv u^\mu\partial_\mu$, $\Theta\equiv\partial_{\mu}\,u^\mu$, $\nabla_{\alpha}=\Delta_{\mu\alpha}\partial^{\mu}$ 
and $A_{(\mu}\,B_{\nu)}=\frac{1}{2}
[A_\mu\,B_\nu+A_\nu\,B_\mu]$ gives the symmetrization. \\

We employ Bjorken's prescription\cite{bjor} to describe the one dimensional boost invariant expanding flow, 
were we use the convenient parametrization of the coordinates using the proper time $\tau = \sqrt{t^2-z^2}$ and space-time 
rapidity $y=\frac{1}{2}\,ln[\frac{t+z}{t-z}]$; $t=\tau$ cosh$\,y$ and $z=\tau$ sinh$\,y$. Then the four velocity is given by,
\begin{equation}
u^\mu=(\rm{cosh}\,y,0,0,\rm{sinh}\,y).
\end{equation}
We note that with this transformation of the coordinates, $D=\frac{\partial\,}{\partial\tau}$ and $\Theta=1/\tau$. 

Form of the energy momentum tensor in the local rest frame 
of the fireball is then given by\cite{Teaney03,SHC06,AM07,AM-prl02}: 
\begin{equation}
T^{\mu\nu} = \left(
\begin{array}{cccc}
\varepsilon & 0 & 0 & 0 \\
0 & P_{\perp} & 0 & 0 \\
0 & 0 & P_{\perp} & 0 \\
0 & 0 & 0 & P_{z} 
\end{array} \right) 
\label{Tmunu}
\end{equation}
where the effective pressure of the expanding fluid 
in the transverse and longitudinal directions are respectively given by 
\begin{eqnarray}
P_{\perp} &=& P + \Pi + \frac{1}{2}  \Phi
\nonumber \\
P_{z} &=& P + \Pi - \Phi
\label{pressures}
\end{eqnarray}
Here $\Phi$ and $\Pi$ are the non-equilibrium contributions to the  equilibrium pressure $P$
coming from shear and bulk viscosities respectively. Respecting the symmetries in the 
transverse directions the traceless
shear tensor has the form $\pi^{ij} = \mathrm{diag}(\Phi/2, \Phi/2,-\Phi)$.

In the first order Navier-Stokes dissipative hydrodynamics 
\begin{equation}
  \label{NS}
  \Pi = -\zeta \partial_\mu u^\mu \quad \rm{and} \quad \pi^{\mu\nu} = \eta 
   \nabla^{\langle\mu} u^{\nu\rangle} \, ,
\end{equation}
with $\zeta,\eta>0$ and $\nabla_{\langle\mu} u_{\nu\rangle}=2\nabla_{(\mu}\,u_{\nu)}
-\frac{2}{3}\,\Delta_{\mu\nu}\nabla_\alpha u^\alpha$. So for first order theories with Bjorken flow we have
\begin{equation}
 \Pi=-\frac{\zeta}{\tau}\quad \rm{and} \quad\Phi=\frac{4\eta}{3\tau}.
\end{equation}
The Navier-Stokes hydrodynamics is known to have
 instabilities and acausal behaviours\cite{lindblom,BRW-06}-- second order theories removes such 
unphysical artifacts.

We use causal dissipative second order hydrodynamics of Isreal-Strewart\cite{Israel:1979wp} to study the 
expanding plasma in the fireball. In this theory we have evolution equations for $\Pi$ and $\Phi$ governed by 
their relaxation times $\tau_{\Pi}$ and $\tau_{\pi}$. 
We refer \cite{AM-DR:04,ROM:09R} for more details on the recent developments 
in the theory and its application to relativistic heavy ion collisions.

Under these assumptions, the set of equations (i.e., equation of motion (\ref{edot}) and relaxation equations 
for viscous terms) dictating the longitudinal expansion of the medium are given by\cite{AM07,BRW-06,Heinz05} 
\begin{eqnarray}
   \frac{\partial\varepsilon}{\partial\tau} &=& - \frac{1}{\tau}(\varepsilon 
  + P +\Pi - \Phi) \, ,
  \label{evo1} \\
  \frac{\partial\Phi}{\partial\tau} &=& - \frac{\Phi}{\tau_{\pi}}+\frac{2}{3}\frac{1}{\beta_2\tau}
  -\left[ \frac{4\tau_{\pi}}{3\tau}\Phi +\frac{\lambda_1}{2\eta^2}\Phi^2\right] 
  \, , 
  \label{evo2} \\
   \frac{\partial\Pi}{\partial\tau} &=& - \frac{\Pi}{\tau_{\Pi}} - \frac{1}{\beta_0\tau} .
\label{evo3}
\end{eqnarray}
where $\Phi=\pi^{00}-\pi^{zz}$.
The terms in the square bracket in Equation(\ref{evo2}) 
are needed for the conformality of the 
theory\cite{Baier:2008:JHEP}. The coefficients $\beta_0$ and $\beta_2$ are related with the relaxation time by
\begin{equation}
 \tau_\Pi=\zeta\,\beta_0\,,\tau_\pi=2\eta\,\beta_2.
\end{equation}
We use the $\mathcal N\,=\,4$ supersymmetric Yang-Mills theory expressions for $\tau_\pi$ and $\lambda_1$
\cite{Baier:2008:JHEP,Okamura:N=4,shiraz:N=4}:
\begin{equation}\label{tau_pi}
  \tau_{\pi} = \frac{2-\ln 2}{2\pi T} 
\end{equation}
and 
\begin{equation}
 \lambda_1 = \frac{\eta}{2\pi T} \, .\label{lambda}
\end{equation}
We set $\tau_\pi(T)=\tau_\Pi(T)$ as we don't have any reliable prediction for $\tau_\Pi$\cite{fms08}.

In order to close the hydrodynamical evolution equations (\ref{evo1} - \ref{evo3}) we need to supply the EoS.

\subsection{Equation of state, $\zeta/s$ and $\eta/s$}

We are interested in the effect of bulk viscosity on the hydrodynamical evolution of the plasma 
and recent studies show that near the critical temperature $T_c$ effect of bulk viscosity becomes 
important\cite{karsch07,Kharzeev:2007wb,GDMoore08,ROM-SON09,caron09}. 
We use the recent lattice QCD result of A. Bazavov $\it{et ~al.}$\cite{Bazavov:2009} for equilibrium equation of state 
(EoS) (\textit{non-ideal}: $\varepsilon-3P\neq 0$). 
Parametrised form of their result for trace anomaly is given by
\begin{equation}
\frac{\varepsilon-3P}{T^4} = \left(1-\frac{1}{\left[1+\exp\left(\frac{T-c_1}{c_2}\right)\right]^2}\right)
\left(\frac{d_2}{T^2}+\frac{d_4}{T^4}\right)\ ,
\label{e3pt4}
\end{equation}
where values of the coefficients are $d_2= 0.24$~GeV$^2$, $d_4=0.0054$~GeV$^4$, $c_1=0.2073$~GeV, 
and $c_2=0.0172$~GeV\cite{kr:2010cv}. Their calculations predict a crossover from QGP to hadron gas around 0.2-0.18 GeV. 
We take critical temperature $T_c$ as 0.19 GeV throughout the analysis. 
The functional form of the pressure is given by \cite{Bazavov:2009}
\begin{equation}
\frac{P(T)}{T^4} - \frac{P(T_0)}{T_0^4} = \int_{T_0}^T dT' \,\frac{\varepsilon-3P}{T'^5}\ ,
\label{pt4}
\end{equation}
with $T_0~$= 50 MeV and $P(T_0)$ = 0 \cite{kr:2010cv}.

From Equations (\ref{e3pt4}) and (\ref{pt4}) we get $\varepsilon$ and $P$ in terms of $T$.

\begin{figure}
\includegraphics[width=8.5cm,height=6cm]{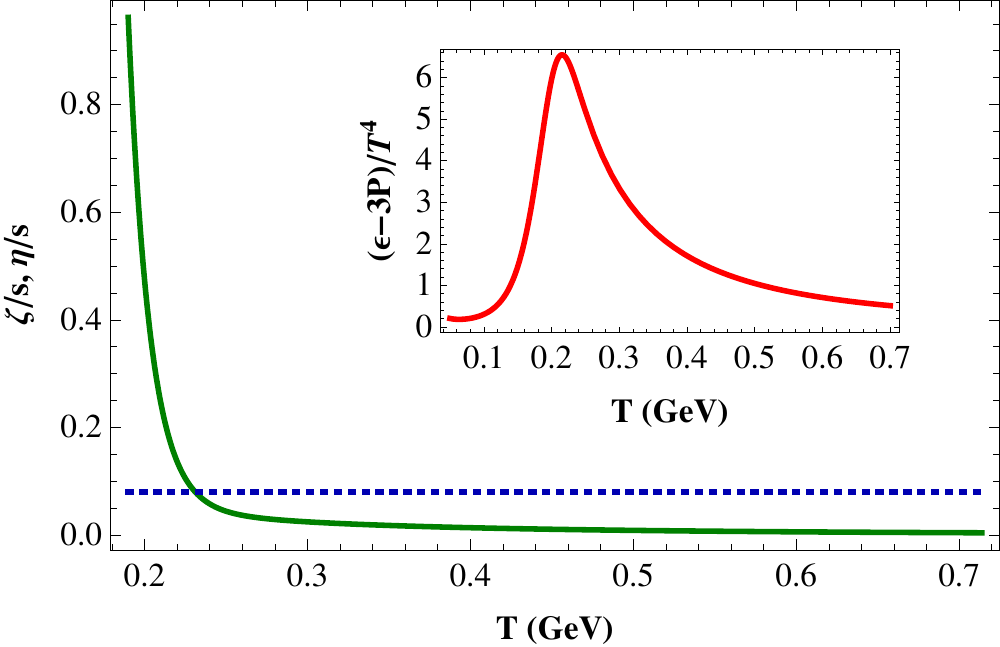}
\caption{$(\varepsilon-3P)/T^4\, ,\zeta/s$ (and $\eta/s=1/4\pi$) as functions of temperature T. 
One can see around critical temperature ($T_c=.190$ GeV)  $\zeta\gg\eta$ and departure of equation of state 
from ideal case is large.}
\label{fig1}
\end{figure}

We rely upon the lattice QCD calculation results for determining $\zeta/s$. We use the result of 
Meyer\cite{Meyer:2008}, which indicate the existence a peak of $\zeta/s$ near $T_c$, however the height 
and width of this curve are not well understood. 
We follow parametrization of Meyer's 
result from Ref.\cite{kr:2010cv}, given by
\begin{equation}
\frac{\zeta}{s} = a \exp\left( \frac{T_c - T}{\Delta T} \right) + b \left(\frac{T_c}{T}\right)^2\quad{\rm for}\ T>T_c,
\label{zetabys}
\end{equation}
where $b$ = 0.061. The parameter $a$ controls the height and $\Delta T$ controls the width of the $\zeta/s$ curve 
and are given by
\begin{equation}
a=0.901,\, \Delta T=\frac{T_c}{14.5} .\label{zetacurve}
\end{equation}
We will change these values to explore the various cases of $\zeta/s$ to account for the uncertainty of the height 
and width of the curve.

\begin{figure}
\includegraphics[width=8.5cm,height=6cm]{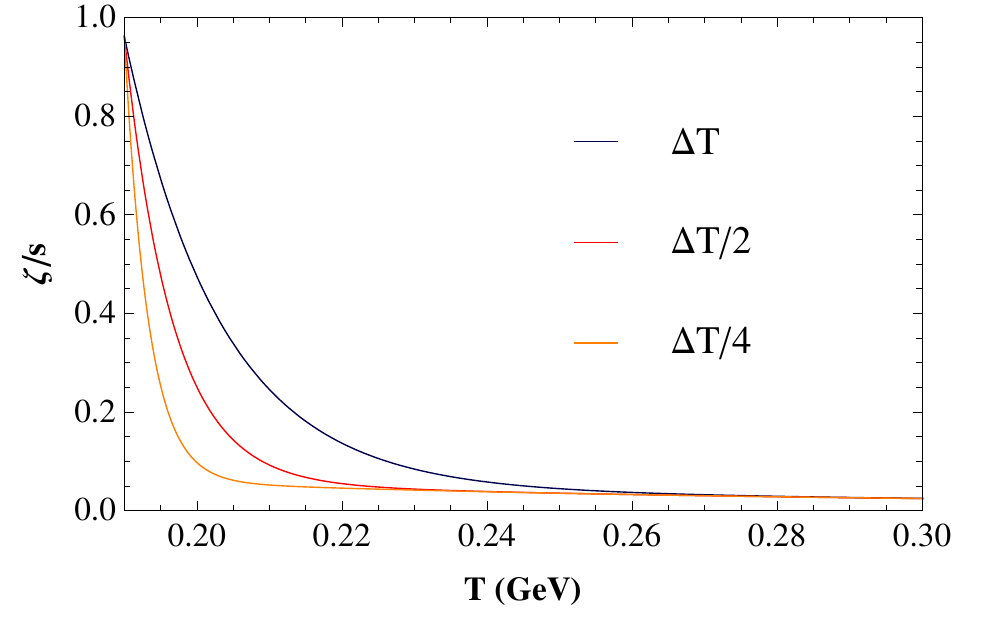}
\caption{Various bulk viscosity scenarios by changing the width of the curve through the parameter $\Delta T$.}
\label{fig1a}
\end{figure}

We use the lower bound of the shear viscosity to entropy density ratio known as KSS bound\cite{kss05} 
\begin{equation}
\eta/s=1/4\pi \label{KSS}
\end{equation}
in our calculations. We note that the entropy density is obtained from the relation
\begin{equation}
 s=\frac{\varepsilon+P}{T}\label{s}.
\end{equation}

In Fig.[\ref{fig1}] we plot the trace anomaly $(\varepsilon-3P)/T^4$ and $\zeta/s$ for desired temperature range. 
We also plot the constant value of $\eta/s=1/4\pi$ for a comparison. It is clear that the \textit{non-ideal} 
EoS deviates from the \textit{ideal} case ($\varepsilon=3P$) significantly around the critical temperature. 
Around same temperature $\zeta/s$ starts to dominate over $\eta/s$ significantly. We would like to note that these 
results are qualitatively in agreement with Ref.\cite{fms08}.

In Fig.[\ref{fig1a}] we show the change in bulk viscosity profile by varying the width of the 
$\zeta/s$ curve by keeping the height intact.

\subsection{Thermal photons}
\label{photons}

During QGP phase thermal photons are 
originated from various sources, like \textit{Compton scattering} $q(\bar q)g\rightarrow q(\bar q)\gamma$ and annihilation 
processes $q\bar q\rightarrow g\gamma$. 
Recently Aurenche \textit{et al.} showed that two 
loop level \textit{bremsstrahlung} process contribution to photon production is as important as \textit{Compton} or 
\textit{annihilation} contributions evaluated up to one loop level\cite{aurenche:98}. They also discussed a new mechanism for hard 
photon production through the annihilation of an off-mass shell quark and an antiquark, 
with the off-mass shell quark coming from scattering 
with another quark or gluon. 
These processes in the context of hydrodynamics of heavy ion collisions were studied in 
Refs.\cite{dk99:EPJC,thoma07}. Until recently only the processes of 
\textit{Compton scattering} and \textit{$q\bar q$-annihilation} were considered in studying the photon production 
rates.

The production rate for hard ($E > T$) thermal photons from 
equilibrated QGP evaluated to the one loop order using perturbartive thermal QCD based on hard thermal loop (HTL) 
resummation to account medium effects. The \textit{Compton scattering} and \textit{$q\bar q$-annihilation} 
contribution to the photon production rate is\cite{kapu91,baier92,traxler95}
 
\begin{equation}
 E\frac{dN}{d^4xd^3p}~=~\frac{1}{2\pi^2}\alpha\alpha_s \left(\sum_f e_f^2\right)~T^2~e^{-E/T}~\rm{ln}
\left(\frac{cE}{\alpha_s T}\right),\label{Compt+Ann}
\end{equation}
where the constant $c\approx$ 0.23 and $\alpha$ and $\alpha_s$ are the  electromagnetic and strong 
coupling constants respectively. In the summation $f$ is over the flavours of the quarks and $e_f$ is the electric charge of the 
quark in units of the charge of the electron.

The rate of photon production due to \textit{Bremsstrahlung} processes is given by\cite{aurenche:98}
\begin{equation}\label{Brems}
 E\frac{dN}{d^4xd^3p}~=~\frac{8}{\pi^5}\alpha\alpha_s \left(\sum_f e_f^2\right)~\frac{T^4}{E^2}~e^{-E/T}~(J_T-J_L)~I(E,T),
\end{equation}
where $J_T\approx1.11$ and $J_L\approx−1.06$ for two flavours and three colors of quarks\cite{thoma07}. 
The expression for $I(E,T)$ is given by
\begin{widetext}
\begin{equation}
 I(E,T)=\left[ 3\zeta(3) +\frac{\pi^2}{6}\frac{E}{T}+\left(\frac{E}{T}\right)^2 \rm{ln}(2) 
        + 4~\rm{Li}_3\left(-e^{-|E|/T}\right) + 2\left(\frac{E}{T}\right)\rm{Li}_2\left(-e^{-|E|/T}\right)
        - \left(\frac{E}{T}\right)^2~\rm{ln}\left(1+e^{-|E|/T}\right)\right],
\end{equation}
\end{widetext}
 
where $\rm{Li}$ are the polylogarithmic functions given by
\begin{equation*}
\rm{Li_a}(z)=\sum_{n=1}^{+\infty}\frac{z^n}{n^a}.
\end{equation*}
Next the rate due to $q\bar q$\textit{-annihilation with an additional scattering in the medium} is given by,
\begin{equation}\label{A+S}
 E\frac{dN}{d^4xd^3p}~=~\frac{8}{3\pi^5}\alpha\alpha_s \left(\sum_f e_f^2\right)~E~T~e^{-E/T}~(J_T-J_L).
\end{equation}

We use the parametrization of $\alpha_s(T)$ by Karsch\cite{karsch88}:
\begin{equation}
 \alpha_s(T)=\frac{6\pi}{(33-2N_f)~\rm{ln(8T/T_c)}}
\end{equation}
for our rate calculations. Here $N_f$ is the number of quark flavors in consideration.\\

\begin{figure}
\includegraphics[width=8.5cm,height=6cm]{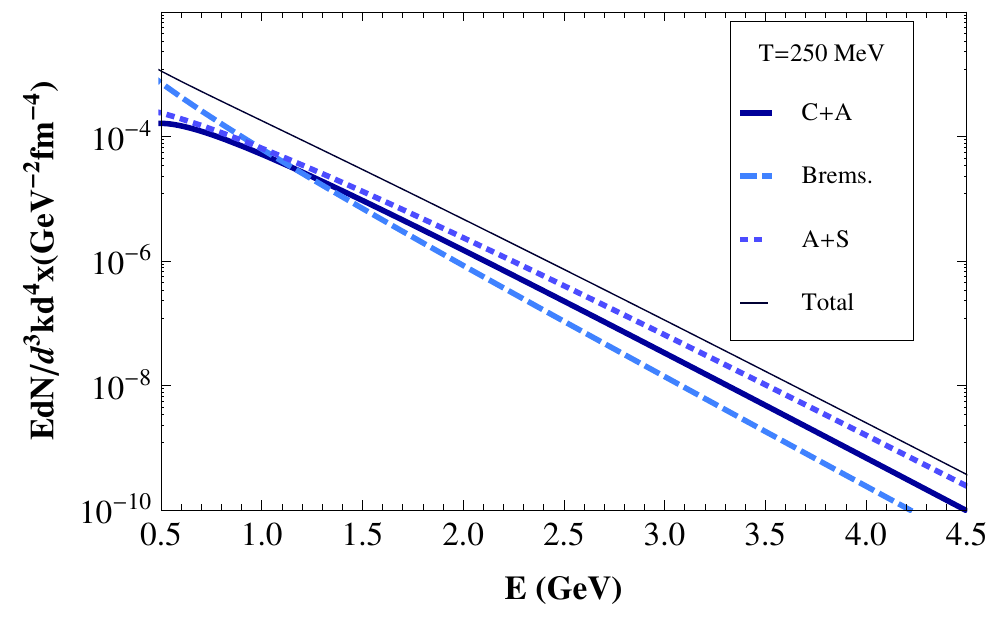}
\caption{Hard thermal photon rates in QGP as a function of energy for a fixed temperature T=250 MeV. Photon rates are 
plotted for different relevant processes.}
\label{fig2}
\end{figure}
In Fig.[\ref{fig2}], we plot the different photon rates for a fixed temperature $T=250~MeV$. It shows the contributions from 
\textit{Bremsstrahlung} (Brems), \textit{annihilation with scattering} (A+S) and \textit{Compton scattering} together with 
\textit{$q\bar q$-annihilation} (C+A). \textit{Bremsstrahlung} contributes to the photon production rate 
upto $E\sim 1~GeV$ only, afterwards A+S and C+A processes become dominant. We might mention here that this observation is 
in agreement with Ref.\cite{thoma07}.\\

The total photon rate is obtained by adding different temperature depended photon rate expressions.
Once the evolution of temperature is known from the hydrodynamical model, the \textit{total photon spectrum} 
is obtained by integrating the total rate over the space time history of the collision\cite{thoma},
\begin{eqnarray}\label{tot-yield}
\left(\frac{dN}{d^2p_T dy}\right)_{y,p_T}&=&\int d^4x\left(E\frac{dN}{d^3pd^4x}\right)\\
&=&Q\int_{\tau_0}^{\tau_1}
d\tau ~\tau \int_{-y_{nuc}}^{y_{nuc}}dy^{'}
\left(E\frac{dN}{d^3pd^4x}\right)\nonumber
\end{eqnarray}
where $\tau_0$ and $\tau_1$ are the initial and final values of time we are interested. 
$y_{nuc}$ is the rapidity of the nuclei whereas $Q$ is its transverse cross-section. 
For a $Au$ nucleus $Q \sim 180 fm^2$. $p_T$ is the photon momentum
in direction perpendicular to the collision axis.
The quantity $\left(E\frac{dN}{d^3pd^4x}\right)$ is  Lorentz invariant
and it is evaluated in the local rest frame in equation (\ref{tot-yield}). 
Now the photon energy in this frame, i.e., in the frame comoving with
the plasma, is given as $p_T cosh(y-y^\prime)$. So once the rapidity and $p_T$ are given we get the 
total photon spectrum.

\subsection{Viscous corrections to the distribution functions}

Viscous effects contribute in two ways in kinetic theory: Firstly, it can change the width (temperature) of the distribution 
function. Secondly, it can modify the momentum dependence of the distribution function. The first effect is incorporated when we 
calculate the temperature as a function of time using dissipative hydrodynamics. To include the second effect one needs to 
compute the change in the distribution function as a function of momentum using the techniques of kinetic theory\cite{deGroot}. 
In the following we give some details of such a calculation.

In section \ref{photons}, while writing the photon rates, we have used Boltzmann distribution function of type 
$f=f_0=e^{-pu/T}$. In order to incorporate the modification due to viscous effects we write the distribution function as 
$f=f_0+\delta f$, with $\delta f=\delta f_\eta + \delta f_\zeta$, where $\delta f_\eta$ and $\delta f_\zeta$ represent change 
in the distribution function due to shear and bulk viscosity respectively. We calculate $\delta f$ using 14-moment Grad's 
method. It ought to be mentioned that recent results show that calculation of $\delta f$ using this method fails near freezeout 
region by making $\delta f$ even larger than $f_0$ and $f<0$ \cite{sasaki09,koide09,piotr10}. It is therfore important to here 
that we are applying these corrections to calculate the photon production rate of hard thermal photons in the regime $T>T_c$. 
We have found that for $p_T$ below 3 GeV, this approximation is reasonable but beyond it, this approximation 
breaks down as the contribution arising from the viscous correction $\delta f$ to the distribution function becomes larger 
than $f_0$\cite{Monnai-Hirano09}. With this caveat we proceed to calculate 
$\delta f$ applying the techniques used in Refs.\cite{Teaney03,DT08}.

We write the viscous correction to the (Boltzmann) distribution function as
\begin{eqnarray} 
f(p) &=& f_0+\delta f = f_0+\delta f_\eta+\delta f_\zeta \\ \nonumber
&=& f_0\bigg(1 + \frac{C}{2 T^3} p^{\alpha}p^{\beta} 
    \nabla_{\langle\alpha} u_{\beta\rangle} 
+ \frac{A}{2 T^3} p^{\alpha}p^{\beta}\Delta_{\alpha\beta}\Theta\bigg) 
\end{eqnarray}
where we restrict the corrections to $f$ upto quadratic order in momentum. 
In order to find the coefficients $A$ and $C$ we first express 
the energy momentum tensor using $f$,
\begin{eqnarray}
 T^{\mu\nu} &=&  \int \frac{d^3p}{(2 \pi)^3 E} p^{\mu}p^{\nu} f \\ \nonumber
&=&  T^{\mu\nu}_{o} + \eta \nabla^{\langle\mu}u^{\nu\rangle}   + \zeta \Delta^{\mu\nu} \;\Theta,
\end{eqnarray}
so that we have
\begin{eqnarray}
\eta \nabla^{\langle\mu}u^{\nu\rangle} &=& \frac{C}{2 T^3}\left[ \int \frac{d^3p}{(2 \pi)^3 E} p^{\mu}p^{\nu} p^{\alpha} p^{\beta}
       f_{o}\right] \nabla_{\langle\alpha}u_{\beta\rangle},\label{eta-f-c}\\
\zeta \Delta^{\mu\nu} \;\Theta &=& \frac{A}{2 T^3}\left[ \int \frac{d^3p}{(2 \pi)^3 E} p^{\mu}p^{\nu} p^{\alpha} p^{\beta}
       f_{o}\right] \Delta_{\alpha\beta} \Theta.\label{zeta-f-c}
\end{eqnarray}
Now from Eq.[\ref{eta-f-c}] we get the correction $\delta f_\eta$ due to the shear viscosity as given in Ref.\cite{Teaney03} 
by finding out $C$ and we will not repeat that calculation here. Next we will find out the coefficient $A$  by constructing 
a fourth rank symmetric tensor out of $\Delta^{\mu\nu}$ and $u^\mu$ representing the term in square brackets in Eq.[\ref{zeta-f-c}],
\begin{eqnarray}\label{tnsr}
 \frac{A}{2 T^3}\left[ \int \frac{d^3p}{(2 \pi)^3 E} p^{\mu}p^{\nu} p^{\alpha} p^{\beta}
       f_{o}\right]= a_{o}  \left(u^{\mu} u^{\nu} u^{\alpha} u^{\beta}\right) \\ \nonumber
 + a_{1}\left( \Delta^{\mu\nu} u^{\alpha} u^{\beta} + \mbox{permutations} \right) ~~~~~~~~~\\ \nonumber
+ a_{2} \left( \Delta^{\mu\nu} \Delta^{\alpha \beta} + 
\Delta^{\mu\alpha}\Delta^{\nu\beta} + \Delta^{\mu\beta}\Delta^{\nu\alpha} \right) \;.
\end{eqnarray}
Now substituting this expression in Eq. [\ref{zeta-f-c}] and by noting $\Delta_{\mu\nu}\;u^\nu=0\;,~ \Delta_{\mu\nu}\Delta^{\mu\nu}=3$ 
and $\Delta^{\mu\nu}\Delta_{\mu\alpha}=\Delta^\nu_\alpha$ we get $\zeta=5 a_2$. Now by contracting both sides of 
Eq. [\ref{tnsr}] with $\frac{1}{45} \left( \Delta_{\mu\nu} \Delta_{\alpha\beta} + 
\Delta_{\mu\alpha}\Delta_{\nu\beta} + \Delta_{\mu\beta}\Delta_{\nu\alpha} \right)$ we get,
\begin{equation}
 \frac{A}{2 T^3}\int \frac{d^3p}{(2 \pi)^3 E} f_{o} \frac{3}{45}\left[p^2-(u.p)^2\right]^2= a_{2} = \zeta/5.
\end{equation}
Evaluating this expression in the local rest frame of the fluid $u^{\mu}=(1,\vec 0 )$ we get
\begin{equation}
   \zeta = \frac{1}{3} \frac{A}{2T^3} \int \frac{d^3p}{(2\pi)^3 E} 
         f_{o} \, {|{\bf p}|}^{4} \;.
\end{equation}
Now for a Boltzmann gas with $f_0=e^{-pu/T}$ we can calculate the integral and comparing the result with that of the entropy density 
$s$ of an ideal boson gas\cite{Teaney03} we find, $A=\frac{2}{5}\;\zeta/s$. So the viscous correction to the 
distribution function due to both shear \cite{Teaney03} and bulk viscosities are given as
\begin{equation}
 f = f_0\bigg(1 + \frac{\eta/s}{2 T^3} p^{\alpha}p^{\beta} 
    \nabla_{\langle\alpha} u_{\beta\rangle} 
+ \frac{2}{5}\frac{\zeta/s}{2 T^3} p^{\alpha}p^{\beta}\Delta_{\alpha\beta}\Theta\bigg).\label{delF}
\end{equation}

Using the Bjorken's flow one can calculate $\nabla_{\langle\alpha} u_{\beta\rangle}$ and $\Delta_{\alpha\beta}\Theta$ for the 
present problem. Four velocity can be written as $u^\alpha=(cosh~y',0,0,sinh~y')$, where $y'$ is the rapidity. 
Let the four momentum of a particle be parametrised 
as $p^{\alpha}$ = $(m_T coshy,p_T cos\phi_p,p_T sin\phi_p,m_Tcoshy)$, where $m_T^2$ = $p_T^2+m^2$\cite{Teaney03}. 
 
Finally we write the distribution function including viscous correction as
\begin{eqnarray}
f = f_0\bigg(1 &+& \frac{\eta/s}{2 T^3} \left[\frac{2}{3\tau}p_T^2-\frac{4}{3\tau}m_T^2sinh^2(y-y') \right] \\ \nonumber
&-& \frac{2}{5}\frac{\zeta/s}{2 T^3} \left[\frac{p_T^2}{\tau}+\frac{m_T^2}{\tau}sinh^2(y-y') \right]\bigg).\label{delFb}
\end{eqnarray}

This expression for distribution function will be used in estimating the photon production rates in Eq.[\ref{tot-yield}].

\section{RESULTS AND DISCUSSION}

\begin{table}
\caption{Initial conditions for RHIC}
\vskip 0.1 in
\begin{center}
\begin{tabular}{ccc}
\hline

\multicolumn{1}{c}{$y_{nuc}$}&
\multicolumn{1}{c}{$\tau_0$} &
\multicolumn{1}{c}{$\rm{T_0}$} \\
\hline
\hline
$~$   &$(fm/c)$ &$(GeV)$\\
\hline
5.3   &0.5  &.310 \\
\hline
\end{tabular}
\end{center}
\end{table}

In order to understand the temporal evolution of temperature $T(\tau)$, pressure $P(\tau)$ and viscous stresses
- $\Phi(\tau)~\rm{and}~\Pi(\tau)$, we numerically solve the hydrodynamical equations describing the longitudinal expansion 
of the plasma: Eqs.[\ref{evo1}-\ref{evo3}]. We use the \textit{non-ideal} EoS obtained from Eq.[\ref{e3pt4}] and Eq.[\ref{pt4}]. 
Information about viscosity coefficients $\zeta$ and $\eta$ are obtained from Eqs.[\ref{zetabys}-\ref{KSS}] 
using Eq.[\ref{s}]. We need to specify the initial conditions to solve the hydrodynamical equations, 
namely the initial time $\tau_0$ and $T_0$. 
We use the initial values relevant for RHIC experiment given in Table I, taken from Ref.\cite{dk99:EPJC}. 
We will take initial values of viscous contributions as $\Phi(\tau_0)=0$ and $\Pi(\tau_0)=0$. 
We would like to note that our hydrodynamical results are in agreement with that of Ref.\cite{kr:2010cv}.

Once we get the temperature profile we calculate the photon production rates. 
Total photon spectrum $E\frac{dN}{d^3pd^4x}$ (as a function of rapidity, $y$ and transverse momentum 
of photon, $p_T$) is obtained by adding different photon rates using 
Eqs. [\ref{Compt+Ann},\ref{Brems},\ref{A+S}] and convoluting with the space time evolution of 
the heavy-ion collision with Eq.[\ref{tot-yield}]. The final value of time $\tau_1$ is the time at which temperature 
evolves to critical value $\tau_f$, i.e.; $T(\tau_1)=T_c$. In all calculations we will consider the photon 
production in mid-rapidity region ($y=0$) only.

We will be exploring various values of viscosity and its effect on the system. 
Since there is an ambiguity regarding the height and width of $\zeta/s$ curve, we will vary the parameters 
$a~\rm{and}~\Delta T$ from its base value given in Eq.[\ref{zetacurve}]. By this we will able to study the effect of 
variation of $\zeta$ on the system. The varied values of the parameters are represented by  $a'~\rm{and}~\Delta T'$. 
We note that unless specified we will be using the base values of bulk viscosity parameters (from Eq.[\ref{zetacurve}]) in our 
calculations. 
Throughout the analysis we will keep the shear viscosity $\eta$ to its base value given by Eq.[\ref{KSS}].

In order to understand the effect of \textit{non-ideal} EoS in hydrodynamical evolution and subsequent photon spectra we 
compare these results with that of an \textit{ideal} EoS ($\varepsilon=3P$).
We consider the EoS of a relativistic gas of massless quarks and gluons. The pressure of such a system is given by
\begin{equation}
P=a~T^4\,;\,a=\left(16+\frac{21}{2}N_f\right)~\frac{\,\pi^2}{90}\label{idealP}
\end{equation}
where $N_f=2$ in our calculations. Hydrodynamical evolution equations of 
such an EoS within ideal (without viscous effects) Bjorken flow can be solved analytically and the 
temperature dependence is given by\cite{bjor}
\begin{equation}
T = T_0~\left(\frac{\tau_0}{\tau}\right)^{1/3} \label{idealT},
\end{equation}
where $\tau_0~\rm{and}~T_0$ are the initial time and temperature.
While considering the viscous effect of this \textit{ideal} EoS, we will solve the set of hydrodynamical equations 
(\ref{evo1} - \ref{evo2}), since effect of bulk viscosity can be neglected in the relativistic limit
when the equation of state $P=\varepsilon/3$ is obeyed \cite{wein}.

\subsection*{Hydrodynamics with \textit{non-ideal} and \textit{ideal} EoS }
\begin{figure}
\includegraphics[width=8.5cm,height=6cm]{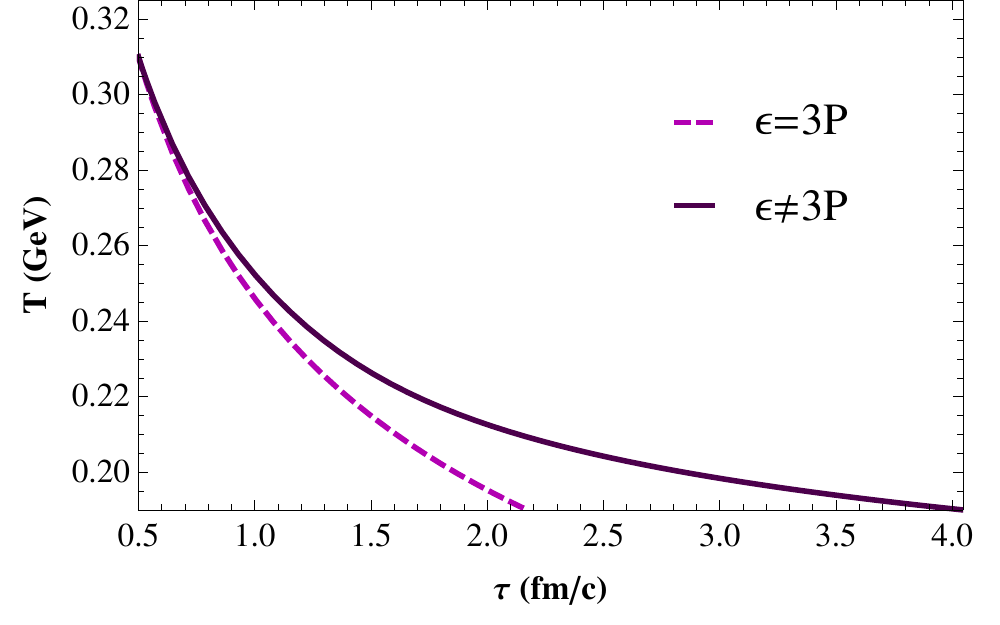}
\caption{Temperature profile using massless (\textit{ideal}) and \textit{non-ideal} EoS in RHIC scenario. 
Viscous effects are neglected in both cases. System evolving with \textit{non-ideal} EoS takes a significantly larger 
time to reach $T_c$ as compared to $ideal$ EoS scenario.}
\label{fig3}
\end{figure}
Fig.[\ref{fig3}] shows plots of temperature versus time for
the \textit{ideal} and \textit{non-ideal} equation of states. The temperature profiles are obtained
from the hydrodynamics without incorporating the effect of viscosity. The figure
shows system with \textit{non-ideal} EoS takes almost the double time than the system with
\textit{ideal} massless EoS to reach $T_c$. So even when the effect of viscosity is not considered, inclusion of 
the \textit{non-ideal} EoS makes significant change in temperature profile of the system. This can affect the corresponding 
photon production rates.

\begin{figure}
\includegraphics[width=8.5cm,height=6cm]{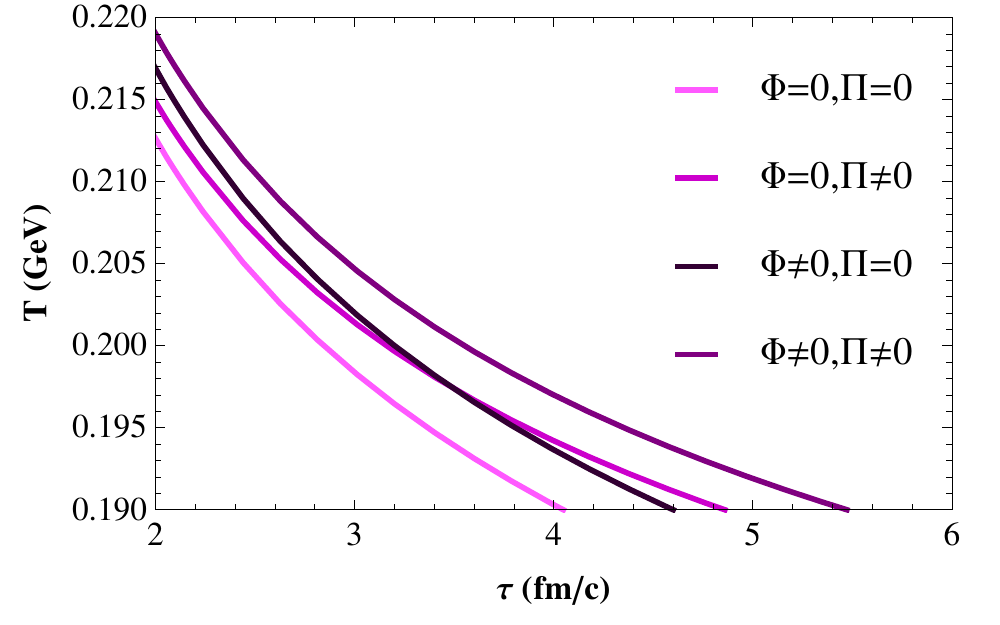}
\caption{Figure shows time evolution of temperature with \textit{non-ideal} EoS for different combinations of bulk 
($\Pi$) and shear ($\Phi$) viscosities. Non zero value of bulk viscosity refers to Eqs.[\ref{zetabys}-\ref{zetacurve}] 
and non zero shear viscosity is calculated from Eq.[\ref{KSS}].}
\label{fig4}
\end{figure}
Next we analyse the viscous effects on the temperature profile. The role of shear viscosity in the boost invariant hydrodynamics 
of heavy ion collisions, for a chemically nonequilibrated system, was already considered in Ref.\cite{skv-photon}. 
We consider possible 
combinations of $\Phi$ and $\Pi$ in \textit{non-ideal} EoS case and study the corresponding temperature 
profiles as shown in Fig.[\ref{fig4}]. As expected viscous effects is slowing down temperature evolution. 
For the case of non zero bulk and shear viscosities ($\Pi\neq 0;~\Phi\neq 0$), temperature takes the longest 
time to reach $T_c$ as indicated by the top most curve. This is about 35$\%$ larger than the case without viscosity (the 
lowest curve). The remaining two curves show that the bulk viscosity dominates over the shear
viscosity when the value of $T$  approaches $T_c$ and this makes the system to spend more time around $T_c$. 
However the intersection point of the two curves may vary 
with values of $a$ and $\Delta T$ as highlighted by Fig.[\ref{fig1a}].

\subsection*{\textit{Non-ideal} EoS and Cavitation}

\begin{figure}
\includegraphics[width=8.5cm,height=6cm]{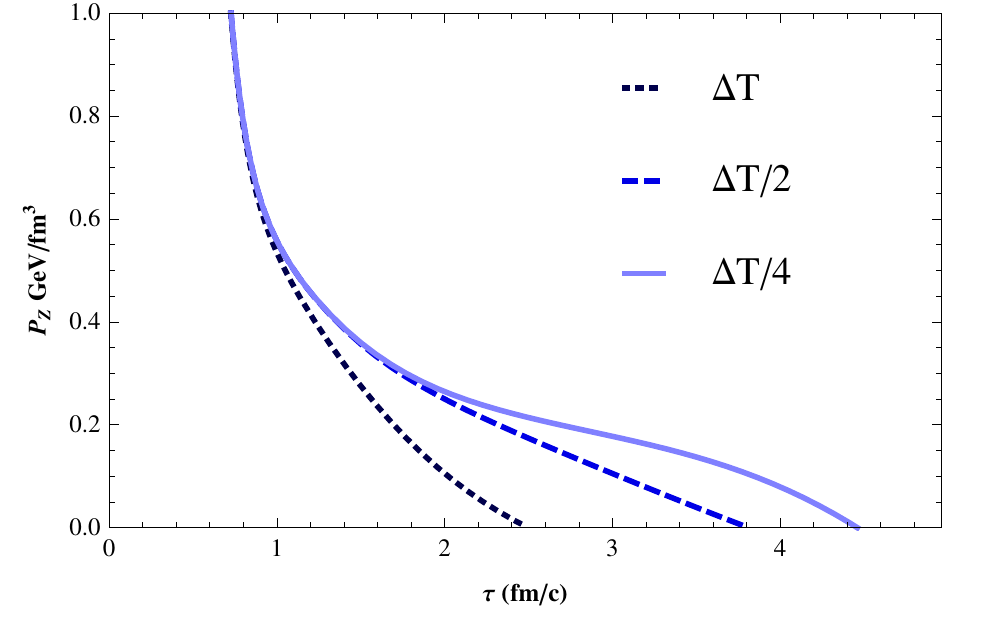}
\caption{Longitudinal pressure $P_z$ for various viscosity cases shown in Fig.[\ref{fig1a}].}
\label{fig5}
\end{figure}

Let us note here the fact that, the bulk viscosity contribution $\Pi$ is negative\cite{kr:2010cv}. 
From the definition of longitudinal pressure $P_{z} = P + \Pi - \Phi$ 
it is clear that if either $\Pi$ or $\Phi$ is large enough it can drive $P_z$ to negative values. 
$Pz=0$ defines the condition for the onset of \textit{cavitation}. During the course of expansion 
when $P_z$ vanishes, the fluid will break apart into fragments and the hydrodynamic treatment will 
become invalid (see for e.g., Ref.\cite{kr:2010cv}). 
Recent experiments at RHIC suggest $\eta/s$ to its smallest value $\sim 1/4\pi$. 
Such a small value of $\eta/s$ alone is inadequate to induce cavitation. Therefore we vary the bulk viscosity values 
by changing $a~\rm{and}~\Delta T$ to study the effect cavitation. In the discussion that follows we will use $\tau_c$ to denote 
the time when cavitation occurs.

\begin{figure}
\includegraphics[width=8.5cm,height=6cm]{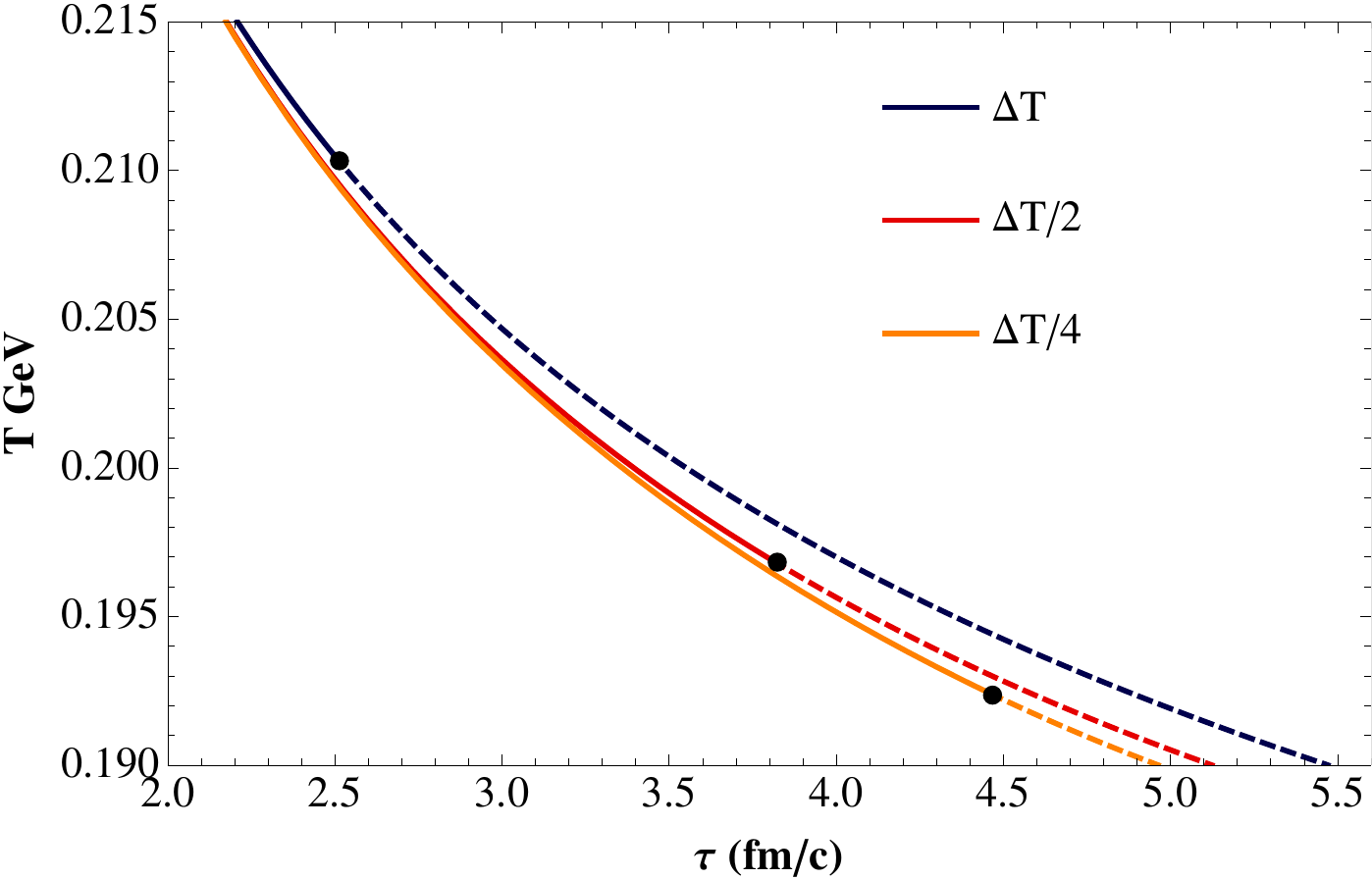}
\caption{Temperature is plotted as a function of time. With peak value ($a$) of $\zeta/s$ remains same while width 
($\Delta T$) varies. In all the three curves, solid lines end at cavitation time $\tau_c$ denoted by a dark circle. 
The dashed lines in each curves show how the system would evolve till $T_c$ if cavitation is ignored. Figure shows 
that larger the width parameter shorter the cavitation time.}
\label{fig7}
\end{figure}

In Figs.[\ref{fig5}-\ref{fig7}] we plot $P_z$ and $T$ as functions of the proper time for different values of $\Delta T$ while 
keeping $a$ (=0.901) fixed. As may be inferred from Fig.[\ref{fig5}], higher value of $\Delta T$ leads to a shorter cavitation time. 
For the values of $a~\rm{and}~\Delta T$ given by Eq.[\ref{zetacurve}] 
we find that around $\tau_c=2.5~fm/c$, $P_z$ becomes zero as shown by the lowest curve in Fig.[\ref{fig5}]. 
In this case, the cavitation occurs when the temperature reaches the value about 210 MeV, as may be seen in Fig.[\ref{fig7}]. 
Had we ignored the cavitation, the system would have taken a time $\tau_f=5.5~fm/c$ to reach $T_c$, which is significantly 
larger than $\tau_c$. This shows that cavitation occurs rather abruptly without giving any sign 
in the temperature profile of the system. The hydrodynamic evolution without implementing the cavitation constraint 
can lead to over-estimation of the evolution time and the photon production.

\begin{figure}
\includegraphics[width=8.5cm,height=6cm]{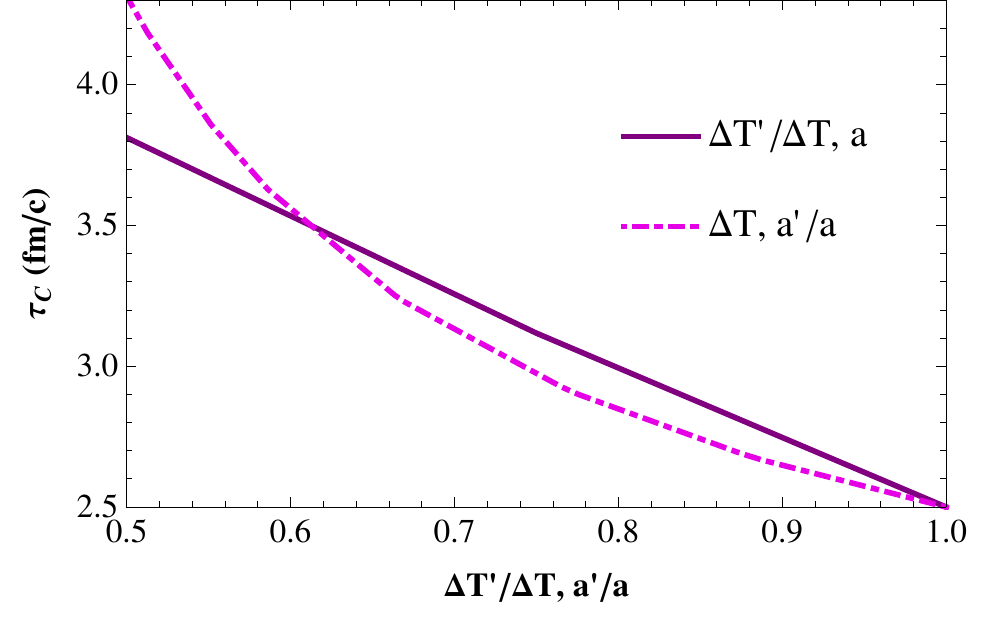}
\caption{Cavitation time $\tau_c$ as a function of different values of 
height ($a'$) and width ($\Delta T'$) of $\zeta/s$ curve.}
\label{fig8}
\end{figure}

We have carried out a similar analysis shown in Figs.[\ref{fig5}-\ref{fig7}] by keeping $\Delta T$ fixed ($=T_c/14.5$) 
and varying parameter $a$. In Fig.[\ref{fig8}] we show the cavitation times corresponding to changes in $a$ and $\Delta T$ 
(denoted by $a'$ and $\Delta T'$). The dashed curve in Fig.[\ref{fig8}] shows $\tau_c$ as a function of $a$, 
while keeping $\Delta T$ fixed. The curve shows that $\tau_c$ decreases with with increasing $a$. 
Solid line shows how $\tau_c$ varies while keeping $a$ fixed and changing $\Delta T$.


\subsection*{Thermal Photon Production}

We have already seen that the calculation of photon production rates require the initial time $\tau_0$, 
final time $\tau_1$ and $T(\tau)$. $\tau_1$ and $T(\tau)$ are determined from the hydrodynamics. Generally $\tau_1$ 
is taken as the time taken by the system to reach $T_c$, i.e.; $\tau_f$. Since hydrodynamics ceases to be valid beyond the 
cavitation time, we must set $\tau_1=\tau_c$. Thus photon production from QGP will be influenced by cavitation, 
temperature profile and \textit{non-ideal} EoS near $T_c$.

\begin{figure}
\includegraphics[width=8.5cm,height=6cm]{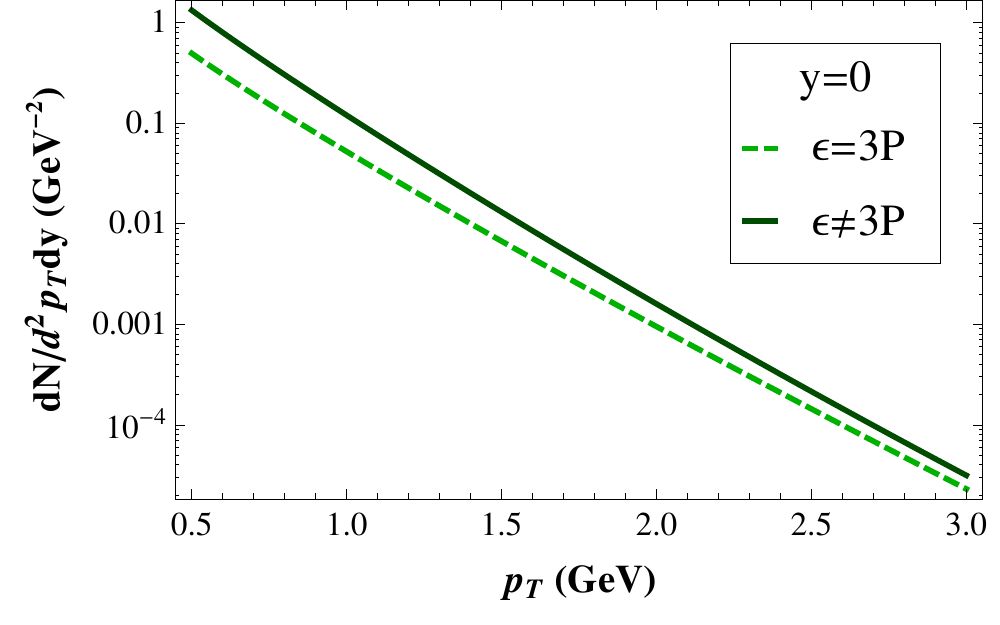}
\caption{Photon flux as function of transverse momentum for different equation of states.
No effect of viscosity included in the hydrodynamical equations and in the distribution functions.}
\label{fig9}
\end{figure}
Fig.[\ref{fig9}] shows the photon production rate calculated using \textit{ideal} (massless) and \textit{non-ideal} EoS. 
The figure shows that \textit{non-ideal} EoS case can yield significantly larger photon flux
as compared to the \textit{ideal} EoS. At $p_T=1$ GeV, photon flux for the \textit{non-ideal} EoS is about 60$\%$ larger 
than that of \textit{ideal} EoS case. This is because the calculation of the
photon flux is done by performing time integral over the interval between the initial time $\tau_0$ and
the final time $\tau_1$. $\tau_0$ is same for both the system while the
$\tau_1$ for the case with \textit{non-ideal} EoS is two times larger than the \textit{ideal} EoS, as may be seen in Fig.[\ref{fig3}]. 
Further, since the \textit{non-ideal} EoS allows the system to have consistently higher
temperature over a longer period as compared to the massless \textit{ideal-gas} EoS, more photons
are produced.

\begin{figure}
\includegraphics[width=8.5cm,height=6cm]{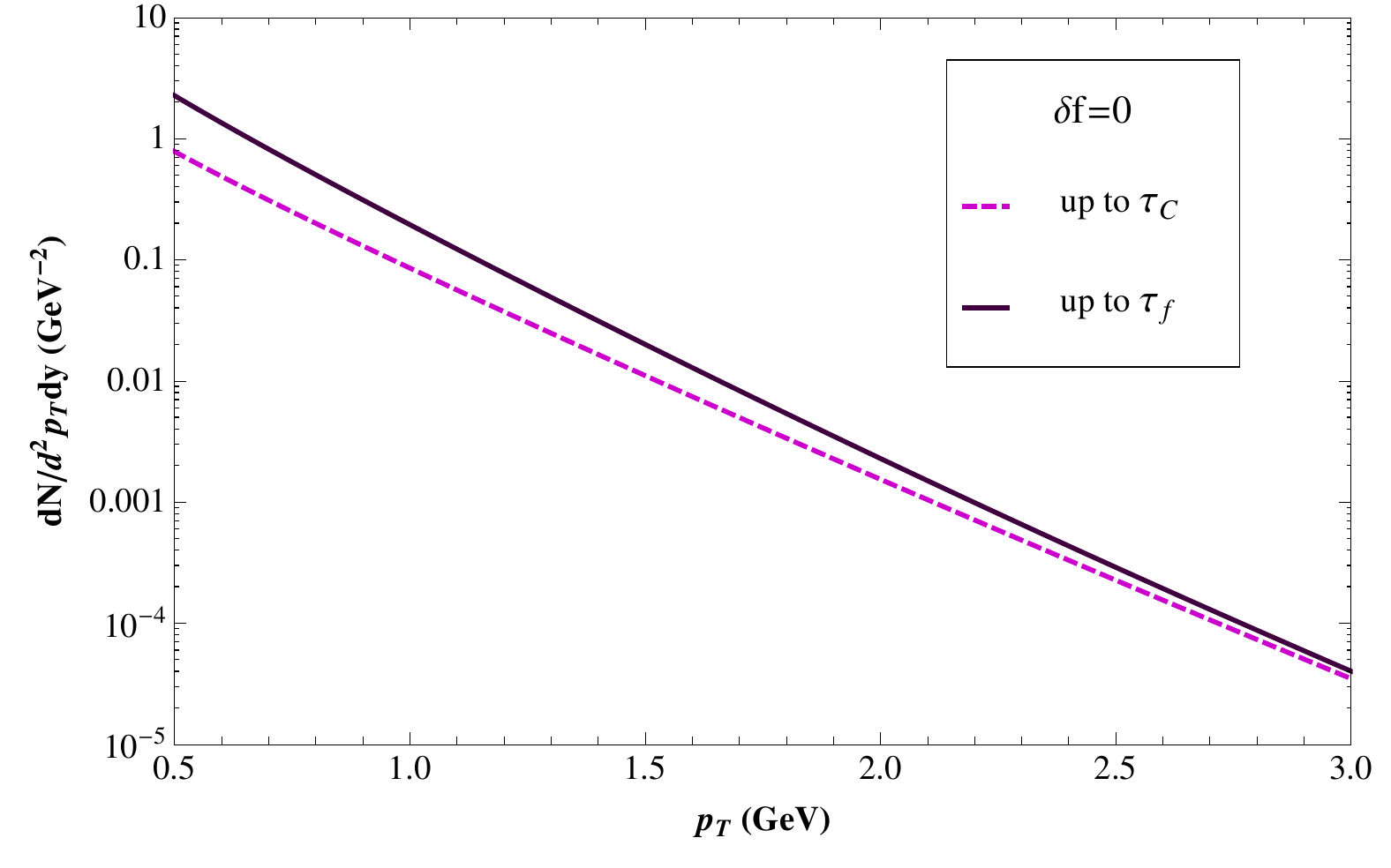}
\caption{Photon spectrum obtained by considering the effect of cavitation (dashed line). For a comparison we plot the 
spectrum without incorporating the effect of cavitation (solid line). Viscous correction to the distribution function 
is not considered.}
\label{fig10a}
\end{figure}

Next, we consider the question that how the cavitation can affect the photon production rate. 
We emphasize that the rates should only be integrated up to the cavitation time $\tau_c$. 
Fig.[\ref{fig10a}] shows the case when there is no viscous correction to the distribution function. In the dashed curve the effect of 
cavitation is taken into account and $\tau_1=\tau_c=2.5~fm/c$. The solid line represents the same case but without the effect 
of the cavitation and $\tau_1=\tau_f=5.5~fm/c$. It can be seen from the curve that ignoring cavitation leads to an over-estimation 
of the rate by about 200$\%$ at $p_T=0.5$ GeV and about 50$\%$ at $p_T=2$ GeV. 
It is thus clear that the information about the cavitation time is crucial for correctly estimating thermal photon production rate.

\begin{figure}
\includegraphics[width=8.5cm,height=6cm]{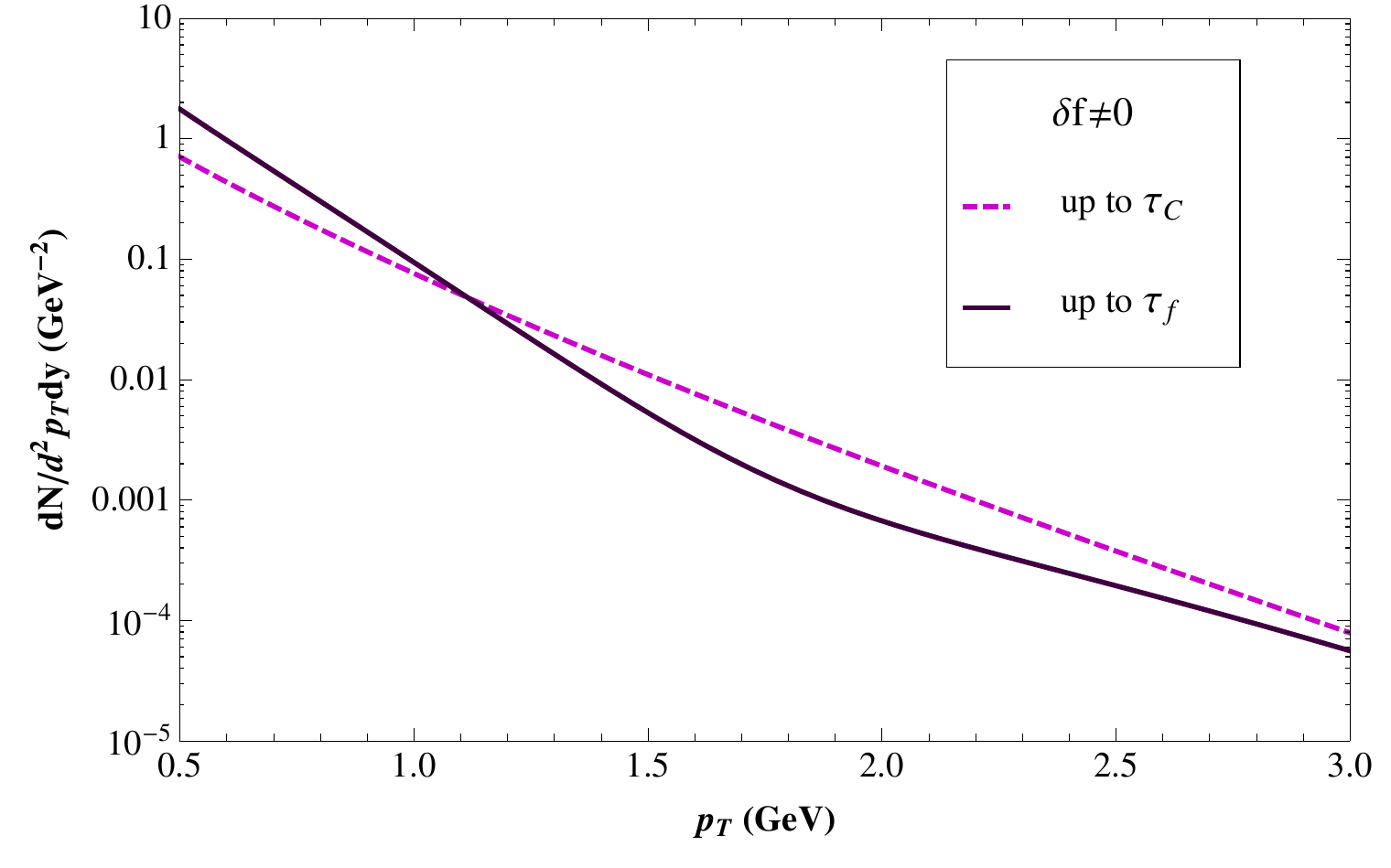}
\caption{Same as in Fig.[\ref{fig10a}] but with incorporating viscous correction to the distribution function.}
\label{fig10b}
\end{figure}

Fig.[\ref{fig10b}] shows the similar comparison between cavitation and no-cavitation cases as in Fig.[\ref{fig10a}], 
but with the inclusion of viscous correction to the distribution function. The solid curve shows the case when cavitation is ignored. 
The dashed curve shows the effect of cavitation. First we note that as can be seen from Fig.[\ref{fig1}], $\zeta/s\gg\eta/s$ 
near $T_c$, while $\eta/s$ dominates over $\zeta/s$ when $T\gg T_c$. If one ignores the cavitation effect, then the 
hydrodynamical code can allow for temperature to evolve upto $T_c$. Moreover, we have also observed that $\delta f_\zeta$ 
contribution dominates over $\delta f_\eta$ contribution for $p_T<$1.5 GeV (A similar behaviour is reported in 
the Ref.\cite{Monnai-Hirano09}). Let us first note that for $\delta f=0$, the photon flux without cavitation is higher as compared 
the same with cavitation at the starting $p_T$ ($\sim 0.5$ GeV). This feature also continues when $\delta f\neq 0$ as can be 
seen in Fig.[\ref{fig10b}]. The negative contribution of $\delta f_\zeta$ on the curve without cavitation (solid curve) makes it 
plummate faster than the curve with cavitation with increasing $p_T$ (dashed curve) and both the curves intersect at $p_T \sim$1.1 
GeV. Beyond this, the shear viscosity starts becoming more effective and prevents this faster plummation of the solid curve. 
Moreover, $\delta f_\zeta$ contribution dominates at $p_T~<1.5$ GeV  Thus in the regime $p_T~<1.5$, ignoring the cavitation effect 
can lead to over-estimation of the photon flux; e.g., at $p_T$ = 0.5 GeV over-estimation is 150$\%$. Whereas there is an 
under-estimation of the photon flux when the effect of cavitation is not included in the high $p_T$ regime. 
E.g., for the following $p_T$ values 1.5, 2.0 and 3.0 (in GeV) under-estimations are around 50$\%$, 65$\%$ and 30$\%$ respectively.

\begin{figure}
\includegraphics[width=8.5cm,height=6cm]{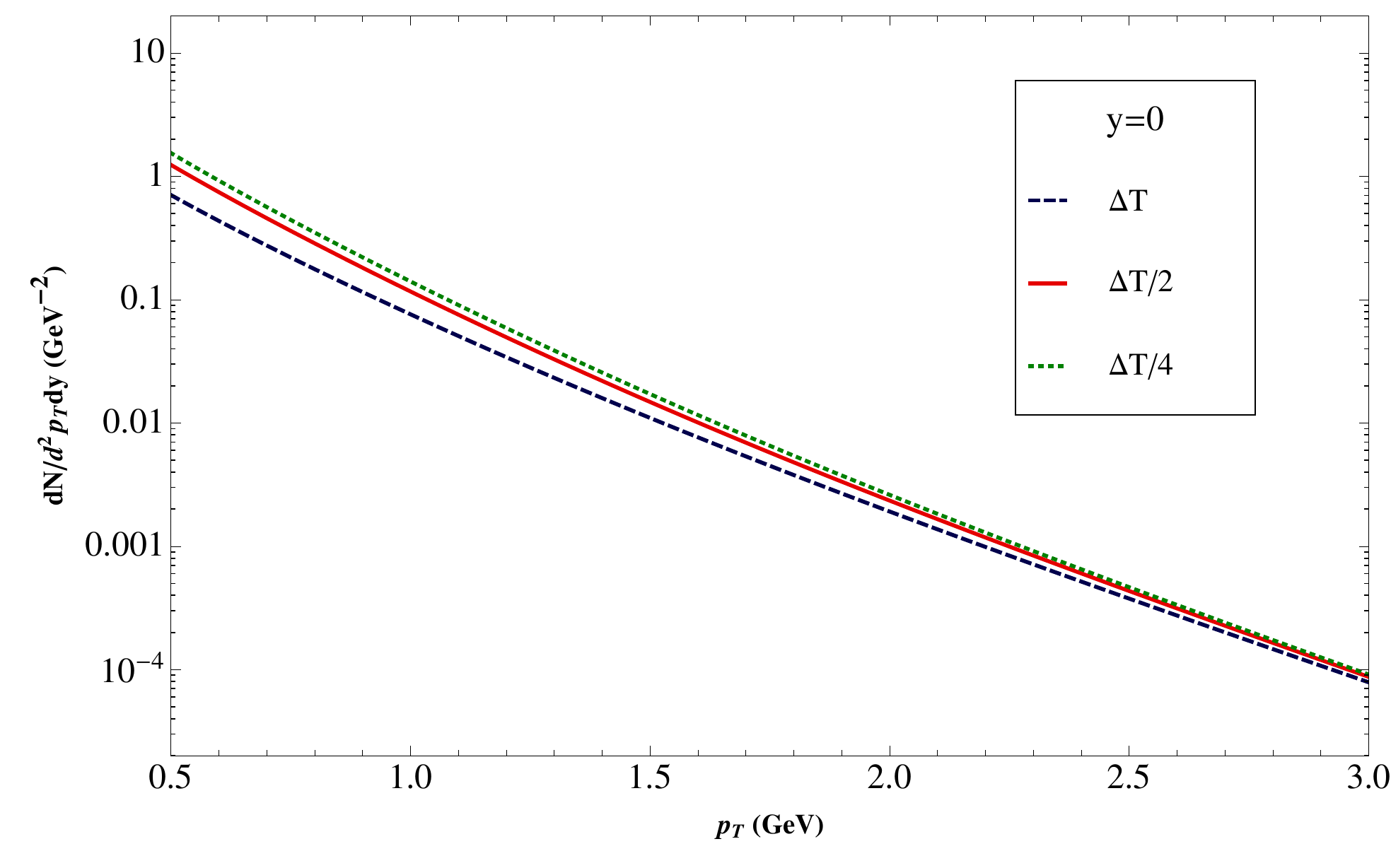}
\caption{Photon production rates showing the effect of different cavitation time. Viscous corrections 
to the distribution functions has been included.}
\label{fig11}
\end{figure}

In Fig.[\ref{fig11}] we plot photon production rates for various cavitation times obtained by varying $\Delta T$ 
(with $a=0.901$ is fixed). Here the enhancement in the photon production when $\Delta T$ is reduced to half of its base 
value is about 75$\%$ at $p_T=0.5$ GeV and about 55$\%$ at $p_T=1$ GeV. A further reduction of the parameter value 
to $\Delta T/4$ is enhancing the photon production by about 120$\%$ at $p_T=0.5$ GeV and about 85$\%$ at $p_T=1$ GeV. 
The reason is a reduction in $\Delta T$ amounts to increase in the cavitation time (see e.g., Fig.[\ref{fig5}]), 
which in turn increases the time interval over which photon production is calculated. 
Therefore this increases the photon flux.

\begin{figure}
\includegraphics[width=8.5cm,height=6cm]{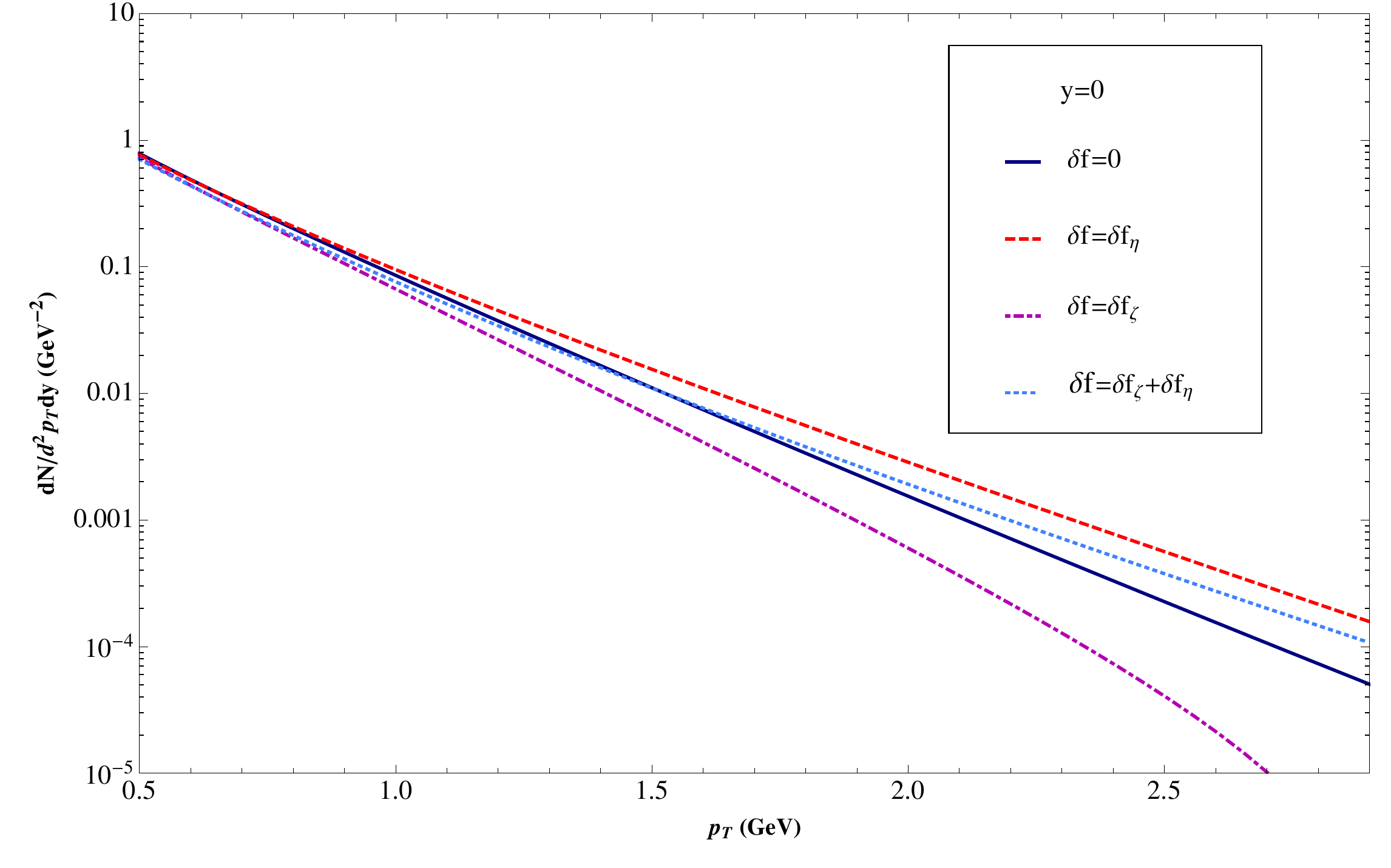}
\caption{Viscous corrections to the distribution function and photon production rate. The solid line shows the 
photon production rate without the viscous corrections to the distribution function and the other lines shows the cases with 
addition of viscous corrections due to shear and bulk viscosities.}
\label{fig12}
\end{figure}

In Fig.[\ref{fig12}] we show the effect of viscous corrections to the distribution function on 
photon production. Here we consider the cavitation scenario with $\Delta T$ ($=T_c/14.5$) and $a$ (=0.901). 
The solid curve shows the case $\delta f=0$. We can see from the figure that when we include only the shear 
viscosity correction ($\delta f=\delta f_\eta$), it enhances the photon production particularly in high $p_T$ 
regime\cite{dusling09}. Next we consider the corrections arising due to bulk viscosity $\delta f=\delta f_\zeta$ only. 
The effective pressure due to $\Pi$ is negative and we have seen that it is crucial for the cavitation to occur 
(See Eq.[\ref{pressures}]). As a result number of particle 
with higher momenta are decreasing and therfore there is a reduction in photon rate as compared to the case without 
the viscous correction\cite{Monnai-Hirano09}. As it is clear from the figure, the effect of bulk viscosity is to oppose 
the contribution from the shear viscosity. The combined effect of both shear and bulk viscosity corrections 
($\delta f=\delta f_\eta+\delta f_\zeta$) can be seen in the graph just below the top-most curve. It is clear from 
the graph that effect of viscous corrections to the distribution functions increases in the high $p_T$ regime.

\section{SUMMARY AND CONCLUSIONS}

Using the second order relativistic hydrodynamics we have analyzed 
the role of non-ideal effects near $T_c$ arising due to the equation of
state, bulk-viscosity and cavitation on the thermal photon production from QGP. 
Since the experiments at RHIC imply extremely small values for $\eta/s$,
we take the value $1/4\pi$ for shear viscosity. 
We have shown using non-ideal EoS from the recent lattice results that 
the hydrodynamical expansion gets significantly slow down as compared
to the case with the massless EoS. This, in turn, enhances the flux of hard 
thermal photons. 

Bulk viscosity plays a dual role in heavy-ion collisions: On one hand it enhances
the time by which the system attains the critical temperature, while on the
other hand it can make the hydrodynamical treatment invalid much before
it reaches $T_c$. Another result we would
like to emphasize is that if the viscous correction $\delta f$ to the distribution function is not included, 
then ignoring cavitation can lead to a significant over-estimation of the photon production rate. But when 
the viscous corrections are included the situation can become more complex. In the low $p_T$ regime ($<$1.1 GeV) 
there is a significant over-estimation in the photon flux if cavitation is ignored. 
On the other hand, in the high $p_T$ regime ($>$1.1 GeV) the photon flux is under-estimated 
when the cavitation is ignored!



\end{document}